\documentclass[twocolumn,showpacs,aps,prl,superscriptaddress]{revtex4}

\usepackage{graphicx}
\usepackage{amsmath}
\usepackage{latexsym}

\begin{document}

\title{Negative Differential Resistance and Astability of the 
Wigner Solid}

\author{G.A. Cs\'athy}
\affiliation{Department of
Electrical Engineering, Princeton University, Princeton, NJ 08544}

\author{D.C. Tsui}
\affiliation{Department of
Electrical Engineering, Princeton University, Princeton, NJ 08544}

\author{L.N. Pfeiffer}
\affiliation{Bell Labs, Lucent Technologies, Murray Hill, NJ 07974}

\author{K.W. West}
\affiliation{Bell Labs, Lucent Technologies, Murray Hill, NJ 07974}

\date{\today}

\begin{abstract}
We report an unusual breakdown of the magnetically induced Wigner solid
in an exceptional two-dimensional electron gas.
The current-voltage characteristic is found to be hysteretic
in the voltage biased setup and has a region of negative differential 
resistance in the current biased setup. When the sample is 
current biased in the region of negative differential resistance,
the voltage on and the current through the sample develop
spontaneous narrow band oscillations. 

\end{abstract}
\maketitle

Two-dimensional electron gases (2DEG) in perpendicular magnetic 
fields $B$ are fascinating systems in which
strong electron-electron interactions lead to the formation of
numerous many body ground states. One well known example is the family of
over 50 fractional quantum Hall liquids \cite{pan}.
Electron solids constitute another class of collective ground states.
The high field insulating phase (HFIP), the reentrant insulating phase (RIP) 
\cite{early,iv,williams,lloyd,lloyd2},
and solids forming close to integer Landau level filling factors 
$\nu$ \cite{chen} have been identified with magnetic field 
induced Wigner solids (WS) \cite{wigner}.
More exotic solids such as the reentrant integer quantum Hall states (RIQHS)
and the electronic liquid crystals of high Landau levels (LL) \cite{lilly}
are believed to be realizations of the theoretically predicted bubble and
stripe phases \cite{koulakov}, respectively. 

The RIQHS shares many properties with the solids enlisted
above \cite{iv,williams,lloyd,lloyd2,chen} 
such as non-linear dc current-voltage characteristics ($I$-$V$) 
with a well defined threshold voltage \cite{jim-iv} and 
sharp resonances in the microwave conductivity \cite{rupert}.
The RIQHS, however, appears to be unique among the 
$B$-field induced solids since it is the only one 
exhibiting spontaneous narrow-band oscillations under dc biasing \cite{jim-osc}.
The nature of these oscillations remains unexplained to date.
Because of the increasing oscillation frequency with an increasing bias 
current \cite{jim-osc}, the oscillations in the RIQHS were suggested to
be akin to the washboard oscillations found in conventional charge
density waves (CDW) \cite{fleming}.
The frequencies of the RIQHS of the order of 1kHz
are, however, orders of magnitude lower than the expected 
washboard frequencies \cite{jim-osc}.

In this Letter we report the observation of spontaneous narrow band oscillations
and of negative differential resistance (NDR) in the WS that forms at the highest
$B$-fields, i.e. the HFIP and the RIP. We used an exceptional
2DEG confined to a GaAs/AlGaAs quantum well and
which has an areal density of $4.8 \times 10^{10}$cm$^{-2}$ 
and a mobility of $13 \times 10^{6}$cm$^2$/Vs.
Unlike the RIQHS, the WS conducts through one single channel
as all electrons occupy the lowest LL. This property 
enables us to construct a simple model that explains our observations. 
We show that the oscillations arise from an astable
behavior of the sample with hysteretic $I$-$V$ connected to the external 
circuitry and that the NDR is a direct consequence of these oscillations.

\begin{figure}[!b]
\begin{center}
\includegraphics[width=2.7in,trim=0.0in 0.0in 0.0in 0.1in]{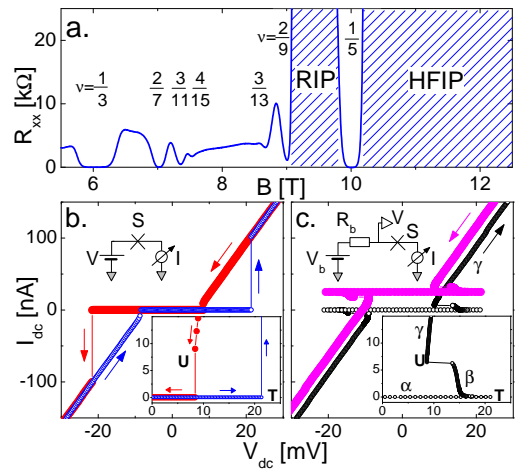}
\end{center}
\caption{\label{f1}
Panel a shows the diagonal resistance $R_{xx}$ as a function of $B$-field 
at 38mK. Lower panels show the $I$-$V$ and the measurement circuitry
in the voltage (panel b) and the current biasing setup (panel c)
when the sample $S$ is at 54mK and at 11T. The curve of
decreasing current (full symbols) of panel c is offseted by 25nA. 
Insets show a magnified view of the low current region.
Branch $\beta$ of panel c exhibits NDR.
}
\end{figure}

Figure.1a. shows the diagonal resistance $R_{xx}$ of our sample at $T=38$mK 
measured in a four wire configuration using a small-signal ac excitation. 
The hashed regions at $\nu<1/5$ and $1/5<\nu<2/9$ mark the HFIP and the RIP.
Resonances in the microwave spectrum recently reported in samples cut
from the same wafer as ours \cite{lloyd2} are considered the most direct
evidence that the RIP and the HFIP is identical to the WS \cite{williams,lloyd,lloyd2}.
We perform two-lead dc $I$-$V$ measurements of the RIP and HFIP. 
As customary in dc measurements, the preamplifiers used are followed
by integrators with their time constant set to 0.3s.
We thus record $I_{dc}$ and $V_{dc}$, the temporal averages of $I$ and $V$. 
Two different biasing configurations are utilized on the same ohmic contacts.
Unless explicitly specified, the sample is held at 54mK and at 11T.

We first employ a voltage biasing configuration shown in Fig.1b.
The $I$-$V$ of the $V$-biased sample
is non-linear and it exhibits a striking hysteretic behavior.
Previously measured $I$-$V$s of the WS 
were also found non-linear \cite{williams,iv} but,
with the exception of one curve from Ref.\cite{williams}, 
none of them showed hysteresis.
Perhaps the most intriguing feature of the $I$-$V$ is its simplicity. 
Indeed, the $I$-$V$ has two linear branches. As the voltage is increased we
find that the current stays virtually zero as long as the voltage
does not exceed the value at the threshold point $T$. 
As the voltage is further increased electrical breakdown occurs when 
the current suddenly jumps to the upper branch that carries non-zero current.
If the $V$ is decreased while the bias point is on the upper branch, 
the current decreases but it jumps to
zero at point $U$ at a voltage lower than that of point $T$.
The current at voltages between that of points $U$ and $T$ is not single valued.

The second measurement setup, shown in Fig.1c, is the current biased 
configuration. The $I$-$V$ obtained has a branch that does not support 
current, one that exhibits an unexpected NDR, 
and one that is at an angle and it is linear to a good approximation. 
These branches are labeled in the inset of Fig.1c by
$\alpha$, $\beta$, and $\gamma$, respectively.
We thus report for the first time NDR in the WS at the highest $B$-fields.
When superimposed (not shown), the linear branches of the two
$I$-$V$s taken with the two different circuits perfectly overlap.
The two $I$-$V$s, however, differ in two ways.
One difference is the NDR is not present in the $V$-biasing setup.
The absence of hysteresis of the $I$-biased configuration,
also shown in Fig.1c, is a second major difference.
The $I$-$V$ of Fig.1c belongs to the so called
S-shaped curves and the NDR is said to be current controlled \cite{sze}.

To understand the origin of the NDR shown in Fig.1c
we explore the time dependence of the voltage on and
current through the sample. We use the same $I$-biased circuit
but we disable the integrators. The bandwidth for $V$ and $I$
is limited by the preamplifiers to 700kHz and 8kHz, respectively.
In Fig.2a we show the $V$-waveforms corresponding to
bias points \#1, \#2, and \#3 of the $I$-$V$ of Fig.2b, an $I$-$V$ which
is identical to that of Fig.1c.
We find that at bias points \#1 and \#3 the
voltage on the sample does not change with time. 
In fact, if the bias point is anywhere on the linear branches $\alpha$ or 
$\gamma$ there is a steady state developed. 
In contrast, when the sample is biased in the NDR region, for example
bias point \#2, there are spontaneous $V$-oscillations present on the sample. 
As seen in Fig.2c, these $V$-oscillations are accompanied by 
synchronized oscillations of the current through the sample. 
Since biasing on the branch of the $I$-$V$ with NDR cannot support
a steady current flow we conclude that this branch has to be understood as
a result of the temporal averaging of the oscillations developed.
Oscillations and NDR are observed
in our sample for all pairs of contacts but only in the $B$-field induced WS,
i.e. the HFIP and RIP shown in Fig.1a,
we therefore think that they are due to an intrinsic bulk effect rather 
than a contact effect.
Using the other biasing configuration, i.e. $V$-biasing,
we detect no oscillations.

\begin{figure}[!t]
\begin{center}
\includegraphics[width=2.7in,trim=0.0in 0.0in 0.0in 0.1in]{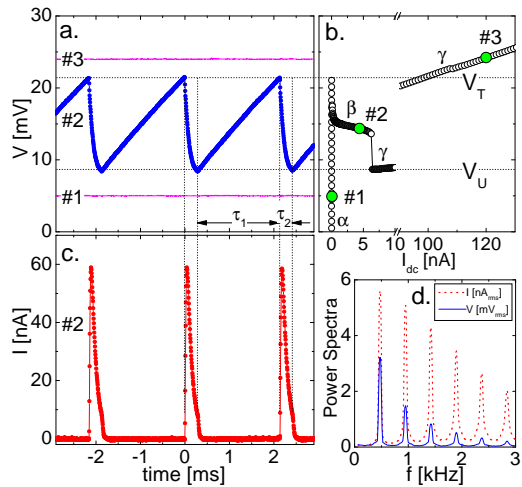}
\end{center}
\caption{\label{f2}
Voltage waveforms (panel a) at three different bias points
of the $I$-biased $I$-$V$ (panel b). Note the break in the $I$-axis
for panel b. At 4.5nA bias (bias point \#2)
both $V$ and $I$ oscillate (panel c). 
Spectra of these oscillations are shown in panel d. 
}
\end{figure}

As seen in Fig.2, the $V$ and $I$ waveforms are quite different but
they have the same frequency and phase.
$V$ has slowly increasing regions that appear to be exponential
and that alternate with quick drops. Correlated with this behavior,
$I$ is zero then it has a narrow spike.
As a consequence, the power spectra of $V$ and $I$
shown in Fig.2d consist of discrete lines at the same frequencies but
have spectral components of different relative weights. 
The quality factor is 400 and in various
cool-downs we measured values as high as 1000. 
In addition a comparison of Fig.2a and Fig.2b reveals that there is
an intricate connection between the oscillations and the $I$-$V$ as
the extremes of the $V$-oscillations coincide with
$V_T$ and $V_U$, the voltages of the endpoints $T$ and $U$ of 
the branches $\alpha$ and $\gamma$ of the $I$-$V$. 

\begin{figure}[!t]
\begin{center}
\includegraphics[width=2.7in]{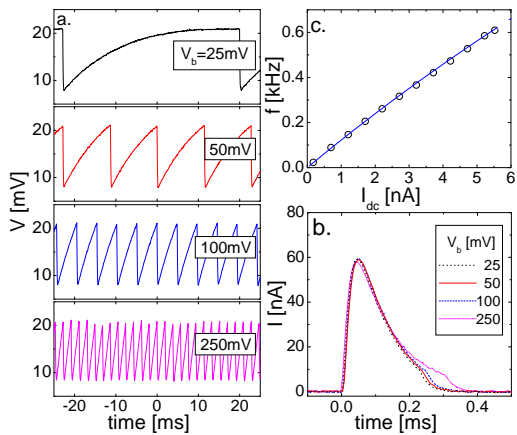}
\end{center}
\caption{\label{f3}
The dependence of the voltage (panel a) and the current oscillations (panel b)
and of the oscillating frequency (panel c) on the biasing conditions.
The line in panel c is the prediction of the model described in the
text and it has no adjustable parameters.
}
\end{figure}

The bias dependence of the oscillations is summarized in Fig.3.
Fig.3a and Fig.3b show that while the amplitude of
both the $V$ and $I$-oscillations does not change with bias, the
oscillation frequency $f$ is strongly bias dependent.
This can better be seen in Fig.3c which displays the
increasing $f$ with the average current through the sample $I_{dc}$.
We found that $f$ is weakly sublinear and it extrapolates to zero as
$I_{dc}$ vanishes. 
The width of the current spikes, shown in Fig.3b,
depends only weakly on the bias. The oscillations are still well developed at
$I_{dc}=5.7$nA but they disappear for $I_{dc}>5.8$nA. In the narrow
current range between the above two values $V$ and $I$ have
burst-like behavior (not shown) that is non-periodic and that is strongly
influenced by fluctuations in the electrical and thermal environment of
the sample. 

We show now that the $I$-biasing circuit of Fig.1c is incomplete. 
Indeed, as seen in Fig.2b,
there are time periods when the current flowing through the sample in zero while
the current through $R_b$ is 4.5nA. 
Kirchhoff's first law of current conservation thus cannot be satisfied unless
an additional circuit element is included. This element is 
the capacitance $C$ of our cables to ground. 
The completed circuit consisting of a non-ohmic
resistor $S$, the capacitance $C$, and the biasing elements $V_b$, $R_b$
is shown in Fig.4. For the model $I$-$V$ of the non-ohmic element
we consider two dc stable branches described by $I=0$ for $V<V_T$ and 
$I=(V-V_0)/r$ for $V>V_U$. From the analysis of this circuit we
obtain that when the 
biasing conditions are such that the load line crosses both branches of the $I$-$V$
(not shown), the circuit is bistable, i.e. depending on the biasing history
the circuit assumes one of the two stable operating points. 
However, when the biasing is such that the load line does not
cross any of the stable branches, shown in Fig.4, the circuit is
astable i.e. it exhibits temporal oscillations between two predetermined states.

\begin{figure}[!h]
\begin{center}
\includegraphics[width=2.7in,trim=0.0in 0.0in 0.0in 0.1in]{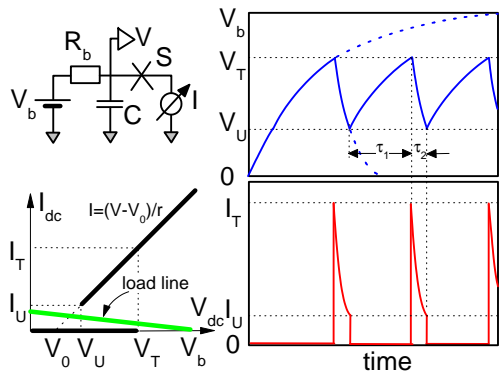}
\end{center}
\caption{\label{f5}
The complete circuit, the model $I$-$V$ of the sample $S$, and
predictions of the model for the time dependences of $V$ and $I$.
}
\end{figure}

We describe now the temporal behavior of the astable circuit.
When we connect the battery to the circuit
the sample does not conduct. As a result the capacitor starts 
charging up towards $V_b$ with the time constant $R_bC$.
When the voltage $V$ on the sample, which is identical to that on $C$, 
reaches its threshold value $V_T$ 
the sample becomes a conductor and the capacitor starts to
discharge through the sample. During a quick discharge the sample current is
bound to take values on the non-zero branch of the $I$-$V$ and
the discharge continues until the voltage on the sample drops to $V_U$.
At this point the charging process begins anew 
leading to cyclical variations of $V$ and $I$ that resemble those of Fig.2.

The simple $I$-$V$ with linear branches allows an exact description
of these processes. After solving the differential equations describing
the motion of the charges
we obtain the charging and discharging time of the capacitor to be
$\tau_1=R_bC \ln (V_b-V_U)/(V_b-V_T)$ and 
$\tau_2=r C \ln ((V_T-V_0)R_b-(V_b-V_0)r)/((V_U-V_0)R_b-(V_b-V_0)r)$,
respectively. However, a simple approximation
will be sufficient to explain our data. 
In the large $R_b$ limit, i.e. for $R_b>>r$ and $V_b>>V_T>V_U$,
the time scales above can be expressed as
\begin{equation}
\tau_1 \simeq C \frac{V_T-V_U}{I_{dc}}, \;\;\;\text{and} \;\;\;
\tau_2 \simeq rC \ln \left( \frac{V_T-V_0}{V_U-V_0} \right). 
\label{eq1}
\end{equation}
The frequency of the oscillations $f=1/(\tau_1+\tau_2)$
is a function of quantities characterizing both the $I$-$V$ of
the sample as well as the external biasing circuitry. 
The parameters $V_T=21.4$mV, $V_U=8.4$mV, 
$V_0=6.5$mV, $r=144$k$\Omega$ extracted from the $I$-$V$ 
and the cable capacitance of $C \simeq 610$pF yield 
$\tau_2=0.18$ms which is close to the width of the
spikes of Fig.3b.
The continuous line on Fig.2c is the prediction of Eq.(\ref{eq1})
for the current dependence of $f$ that has no fitting parameters 
and which is in excellent agreement with our data.

\begin{figure}[h]
\begin{center}
\includegraphics[width=2.7in]{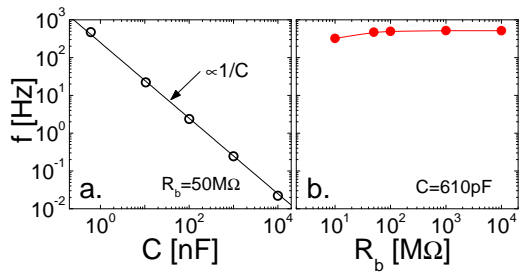}
\end{center}
\caption{\label{f4}
The dependence of the frequency of oscillations on the capacitance to ground $C$
at $R_b$=50M$\Omega$ (panel a) and on the biasing resistor $R_b$ at $C$=610pF 
(panel b) at 4.5nA bias. 
}
\end{figure}

To further test the model we vary the circuit elements $C$ and $R_b$
at a fixed 4.5nA biasing current. As shown in Fig.5a
we find that the frequency at fixed $R_b$ is inversely
proportional to the capacitance $C$, a behavior that is consistent with
the linearly scaling $\tau_1$ and $\tau_2$ with $C$. 
At a given value of $I_{dc}$ and at
fixed value of $C$ in the limit of large $R_b$s
the frequency is found to be independent of $R_b$ as seen in Fig.5b. 
Such a behavior is expected from Eq.\ref{eq1} showing
both $\tau_1$ and $\tau_2$ to be $R_b$-independent. 

According to the described model spontaneous oscillations develop in the 
WS and in devices with a special non-linear $I$-$V$, i.e. 
a hysteretic $I$-$V$ in the $V$-biased setup
and an $I$-$V$ that exhibits NDR in the $I$-biased setup.
These oscillations are known as relaxation oscillations 
and the circuit is called astable multivibrator or Schmitt trigger \cite{schmitt}.
The Shockley diode, thyristor, diac, triac \cite{sze}
and the hot electron diode \cite{hed} are examples of  
semiconductor junction devices with similar $I$-$V$s that have been used
in oscillators. The hysteretic $I$-$V$ of the WS, unlike
that of the devices above, is a genuine bulk effect that can be explained
by the response of the WS to an external bias field in the presence of disorder.
This response is not yet fully understood. In one scenario the WS breaks up into
domains that are pinned by the disorder and the finite conduction results
from motion of the WS domains when the externally applied force overcomes 
the pinning force \cite{fuku}. 
Hysteresis results from the gain of kinetic energy from the external field.
Such an energy gain is forbidden when the WS is pinned but
ensures a flow of charges when the voltage is lowered below its threshold value.
In this interpretation the oscillations arise from the
switching between the pinned and the sliding solid, a process that is similar
to that found in CDW \cite{mihaly}. 
Lattice defects moving through a pinned WS constitute a second scenario.
For example an interstitial electron can move in the periodic potential 
of the other electrons forming the WS. Such a motion 
leads to Bloch energy bands and conduction can arise 
when the external field excites the interstitial electron into the Bloch conduction band. 

To summarize, the $I$-$V$ of the WS forming at the highest $B$-fields
is hysteretic in the $V$-biased setup and
exhibits a region of NDR in the $I$-biasing configuration.
At special biasing conditions spontaneous voltage and current oscillations
are detected that can be accounted for by a simple model. 
This model predicts similar oscillations for the $B=0$ WS in a Si-MOSFET
in which NDR has recently been reported \cite{mos}. Finally we note that
due to the different bias circuits a comparison of the results obtained by us
and those for the RIQHS \cite{jim-iv,jim-osc} 
is not straightforward. Several properties of the two phases, however, 
show resemblance. For both electronic solids the $I$-$V$s are hysteretic, 
oscillations develop in the audio frequency range,
the oscillation frequency increases with increasing bias, and
the oscillation amplitude is independent of the bias. 
This research was funded by the DOE and the NSF.

\end{document}